\documentclass[aps,prl,twocolumn,showpacs]{revtex4}
\usepackage{amsmath,amssymb,array}
\usepackage{graphicx}
\newcommand{\beq}{\begin{eqnarray}}
\newcommand{\eeq}{\end{eqnarray}}

\newcommand{\bvec}[1]{{\mathbf #1}}

\newcommand{\ket}[1]{\left| #1 \right>}

\begin{document}

\title{Enhanced ferromagnetism from electron-electron interactions in 
double exchange type models}

\author{Nuri A. Yazdani and Malcolm P. Kennett }

\address
{Department of Physics, Simon Fraser University, 8888 University Drive, 
Burnaby, British Columbia V5A 1S6, Canada}

\date{\today}
\begin{abstract}
The magnetic
properties of a variety of materials with
promise for technological applications have been described by models
in which fermions are coupled to local moment spins.  Monte Carlo 
studies of such models usually
ignore electron-electron interactions, even though the energy scale 
corresponding to these interactions may be comparable to or larger than 
other relevant energy scales.  In this work we add on-site
interactions between fermions to the double exchange model 
which we study with a Monte Carlo scheme in which
temporal fluctuations of local moment spins are fully accounted for and 
electron-electron interactions
are treated at a mean field level. 
 We show that when the number of fermions is considerably less
than the number of local moments even moderate interactions can 
lead to significant enhancement of ferromagnetism and the 
Curie temperature.
\end{abstract}


\maketitle

The magnetic properties of a variety of materials of fundamental interest
and technological promise, such as collosal magnetoresistance (CMR)
manganites \cite{Dagotto_review}, rare earth hexaborides \cite{CastroNeto1}, 
and diluted, magnetic semiconductors (DMS)
 \cite{Dietl} have been described using
double exchange (DE) type models in which fermions are coupled to local moment 
spins.  In the limit that the number of fermions is considerably less than
the number of local moments, such models generally display ferromagnetism \cite{Anderson}.

In manganites it is well established that electron-electron interactions are
at least as large as the Hund coupling, 
and may be the largest energy scale in the problem 
\cite{Dagotto_review,Satpathy,Ramakrishnan,Golosov}.  The importance of interactions has
also been stressed for DMS \cite{Jungwirth,Fiete,Berciu} and hexaborides 
\cite{CastroNeto2}.  The combination of disorder and electron-electron 
interactions may also play an important role in nanoscale electronic 
phase separation \cite{Shenoy}, which has been argued to be important for
CMR in manganites \cite{Burgy}.  It is hence important 
to develop accurate techniques that 
can account for the effects of electron-electron interactions
in DE type models and to study their effects on magnetic properties of these
models.

Here we introduce a method to study the effects of electron-electron
interactions in DE type models when the energy
scale for interactions is in the experimentally relevant regime of
 no more than a few times the Hund coupling.  We use this method to 
show that even moderate interactions can enhance ferromagnetism and the Curie temperature
$T_c$ significantly when the number of fermions is considerably less than the
number of local moments. 

We combine a Hartree-Fock treatment of electron-electron
interactions with a Monte Carlo scheme for fermions coupled to 
classical local moment spins.  Through the use of exact diagonalizations on
small systems we have determined regions of parameter space where this technique should
be most accurate.  We calculate the
magnetization as a function of temperature for a variety of interaction strengths 
and determine the effects of interactions on the Curie temperature $T_c$.  
Electron-electron interactions can lead to ferromagnetism
in the absence of Hund coupling \cite{Hubbard}, so the enhancement in magnetization 
that we find can be understood as this tendency reinforcing the ferromagnetism that
arises from the Hund coupling.

The general Hamiltonian we consider is of the form

\begin{eqnarray}
{\mathcal H} = {\mathcal H}_{\rm DEM} + {\mathcal H}_{\rm int}, 
\label{eq:interactingDEM}
\end{eqnarray}
with
\begin{eqnarray} 
{\mathcal H}_{\rm DEM} & = & - \sum_{ij} \left[ t_{ij} c_{i\sigma}^\dagger c_{j\sigma} + \,
{\rm h.c.}\right] +
\sum_{ij} J_{ij} \bvec{S}_i \cdot \bvec{s}_i , \nonumber \\
{\mathcal H}_{\rm int} & = & U\sum_i n_{i\uparrow} n_{i\downarrow} , \nonumber
\end{eqnarray}
where $i$ and 
$j$ are site indices, $t_{ij}$ is the hopping integral, $J_{ij}$ is the Hund coupling,
$U$ is the on-site Hubbard repulsion,
$c^\dagger_{i\sigma}$ creates a fermion with spin $\sigma$ on site $i$, $\bvec{S}_i$ is 
a local moment spin on site $i$, and $\bvec{s}_j = \frac{1}{2} \sum_{\alpha,\beta} 
\left(c_{j\alpha}^\dagger\mbox{\boldmath $\sigma$}_{\alpha\beta} c_{j\beta}\right)$ 
is the fermion spin on site $j$.
For simplicity, we assume the hopping is to 
nearest neighbour sites on a cubic lattice, and that the Hund
coupling is purely local: $J_{ij} = J\delta_{ij}$.  We assume that there is a local moment on 
every site in the lattice, and that the fermions have spin-$\frac{1}{2}$ 
(these assumptions can be easily relaxed).

There has been some work 
to study the effects of finite $U$ in models of the form
Eq.~(\ref{eq:interactingDEM})
using exact diagonalizations \cite{Mishra}, DMRG \cite{Malvezzi}
and mean field approximations \cite{Berciu}.  However, these
techniques are not appropriate for studying finite temperature magnetic properties in 
dimensions higher than one taking into account the temporal fluctuations of spins.
In order to determine the finite temperature magnetic properties of the Hamiltonian
Eq.~(\ref{eq:interactingDEM}), it is necessary to perform Monte Carlo simulations.
If there are $N$ local moment spins with spin $S$, then the size of the Hilbert 
space scales as $(2S+1)^N$, hence it is usual to approximate the local moment spin as 
classical, which is often reasonable given that in many systems of interest $S$ is 
larger than $\frac{1}{2}$.  Previous such Monte Carlo simulations 
\cite{Yunoki,Schliemann,Kennett1,Alvarez,Fiete} have restricted their
attention to models of the form Eq.~(\ref{eq:interactingDEM}) with $U = 0$
 with the exception of Ref.~\cite{Fiete}.  In Ref.~\cite{Fiete}
interactions were included for some parameter values using a zero temperature 
variational procedure due to convergence issues with Hartree-Fock and did
not appear to have a strong 
influence on magnetic properties.  The model in Ref.~\cite{Fiete}
also included disorder and spin-orbit coupling, and 
these terms may have influenced the convergence of Hartree-Fock calculations.

We write the classical local moment spins in the form
$\bvec{S}_i = (S_i^z,\phi_i)$, and for a specific arrangement of local moment spins, one
can diagonalize Eq.~(\ref{eq:interactingDEM}) with $U=0$: 
\begin{eqnarray}
{\mathcal H}\left(\{\bvec{S}_i\}\right) =  \sum_m E_m\left(\{\bvec{S}_i\}\right) a^\dagger_m a_m,
\end{eqnarray}
where  $\{E_m\left(\{\bvec{S}_i\right)\}$ are the eigenvalues for a given local moment spin configuration.  $a_m^\dagger$
and $a_m$ are the creation and annihilation operators for the $m^{\rm th}$ eigenstate of $H$:
$$ a_m^\dagger = \sum_{i\sigma} \psi_{im\sigma}c^\dagger_{i\sigma}, \quad a_m = \sum_{i\sigma} \psi^*_{im\sigma} c_{i\sigma}.$$
This allows one to write the fermion free energy as
\begin{eqnarray}
{\mathcal F}(\left\{\bvec{S}_i\right\}) = -\frac{1}{\beta} \sum_{m=1}^{2N}\ln\left(1 + e^{-\beta\left(E_m\left(\{\bvec{S}_i\}\right) - \mu\right)}\right),
\end{eqnarray}
where $\beta = 1/k_B T$ is the inverse temperature and hence the classical partition function for the $N$ classical spins takes the form
\begin{eqnarray}
Z = \left[\prod_{i=1}^N \int_{-1}^1 dS_i^z \int_0^{2\pi} d\phi_i \right] 
e^{-\beta{\mathcal F}(\{\bvec{S}_i\})}.
\end{eqnarray}
Casting the partition function in this form makes it clear that we may use the Metropolis algorithm to determine whether or not
to flip a spin, with the change in the fermion free energy determining whether a spin flip is accepted or rejected.

Introducing a finite $U$ greatly increases the size of the fermion Hilbert space:
for $N$ local moments and $n$ fermions there are 
${(2N)!}/{(2N-n)! n!}$ states, as compared to $2N$ non-interacting states.  This
renders even moderate values of $N$ out of 
reach computationally.   In order to explore larger values of $N$ 
the interaction term must be treated in an approximate fashion, hence our use of 
Hartree-Fock as the simplest self-consistent approach.  The Hartree-Fock
approximation reduces the size of the fermion Hilbert space to $2N$ and allows
for the use of the Monte Carlo scheme outlined above, with Hartree-Fock energies 
replacing the $E_m\left(\{\bvec{S}_i\}\right)$.

We decompose $H_{\rm int}$ in the Hamiltonian Eq.~(\ref{eq:interactingDEM}) as
\begin{eqnarray}
U\sum_i n_{i\uparrow}n_{i\downarrow} & \simeq & U\sum_i\left[\left<n_{i\uparrow}\right>c_{i\downarrow}^\dagger c_{i\downarrow}
 + \left<n_{i\downarrow}\right> c^\dagger_{i\uparrow} c_{i\uparrow} \right. \nonumber \\
& & \left.- \left<c^\dagger_{i\uparrow} c_{i\downarrow}\right>
c_{i\downarrow}^\dagger c_{i\uparrow} - \left<c_{i\downarrow}^\dagger c_{i\uparrow}\right> 
c_{i\uparrow}^\dagger c_{i\downarrow}
\right. \nonumber \\
& & \left.
- \left<n_{i\uparrow}\right>\left<n_{i\downarrow}\right> + \left<c_{i\uparrow}^\dagger c_{i\downarrow}\right>
\left<c_{i\downarrow}^\dagger c_{i\uparrow}\right>\right]   .
\label{eq:HFapprox}
\end{eqnarray}
Approximating $H_{\rm int}$  with Eq.~(\ref{eq:HFapprox}) we can write
$$ H \simeq H_{HF} - U \sum_i \left[\left<n_{i\uparrow}\right>\left<n_{i\downarrow}\right> 
 - \left<c_{i\uparrow}^\dagger c_{i\downarrow}\right> \left<c^\dagger_{i\downarrow} c_{i\uparrow}\right>\right],$$
and the single particle states satisfy
$$ H_{HF} \ket{\phi_m} = \epsilon_m \ket{\phi_m},$$
with single particle energies 
\begin{eqnarray}
\tilde{\epsilon}_m & = & \epsilon_m - \frac{U}{2} \sum_i \left[\left<n_{i\uparrow}\right>\psi^*_{im\uparrow}
\psi_{im\uparrow} + \left<n_{i\downarrow}\right>\psi^*_{im\downarrow} \psi_{im\downarrow} \right. \nonumber \\ & & \left.
 - \left<c_{i\uparrow}^\dagger c_{i\downarrow}\right> \psi^*_{im\downarrow} \psi_{im\uparrow} 
 - \left<c^\dagger_{i\downarrow} c_{i\uparrow}\right> \psi^*_{im\uparrow} \psi_{im\downarrow} \right],
\end{eqnarray}
which we calculate by iterating to self-consistency and 
use to determine the approximate fermion free energy for each local moment spin configuration
$\{\bvec{S}_i\}$.

The Hartree-Fock approximation is uncontrolled, hence in order to ascertain the regions of
parameter space in which the hybrid Hartree-Fock Monte Carlo scheme should be most accurate,
we performed exact diagonalization calculations on 8 site 
$2\times 2\times 2$ systems with 2, 3, and 4 
electrons, and compared the energies of the exact ground state, 
first excited state and second excited state 
averaged over 25 different local moment configurations.  In Fig.~\ref{fig:ED}
we show the relative error $\Delta_E = |E_n - E_n^{HF}|/|E_n|$ 
as a function of $J/t$ and $U/t$
, where $E_n$ is the exact energy and 
$E_n^{HF}$ is the Hartree-Fock approximation to the energy.

Provided $U$ is not too large in comparison to $J$, the Hartree-Fock approximation gives a good account of the low-lying energy levels, and we found 
that for $U \lesssim 3J$ the relative error was less than 5\%
with the error decreasing with decreased $U/J$, hence we restricted
our Monte Carlo simulations to this region.  
We additionally checked the relative error in the densities between the exact 
diagonalization results and the Hartree-Fock approximation and found the best agreement 
in the same region of parameter space as for the energy.

We performed Monte Carlo simulations for cubic systems with $N = 4^3 = 64$
and $N = 6^3 = 216$ local moments, 
with $n=8$ and $n=27$ fermions respectively
corresponding to $n/N = 1/8$ in both cases.
These system sizes are competitive with recent simulations in non-interacting
systems \cite{Alvarez}.  We used a similar equilibration
procedure to Ref.~\cite{Kennett1}, by bringing two replicas of the system into
equilibrium, evolving under Metropolis dynamics, with an additional self-consistent
Hartree-Fock loop as discussed above.
We used the $z$-test \cite{Sprinthall} to determine when there was 95\%
confidence that the two replicas were in equilibrium and then collected data for 
a further 10000 Monte Carlo sweeps.

\begin{widetext}
\begin{center}
\begin{figure}[htb]
\includegraphics[width=12cm]{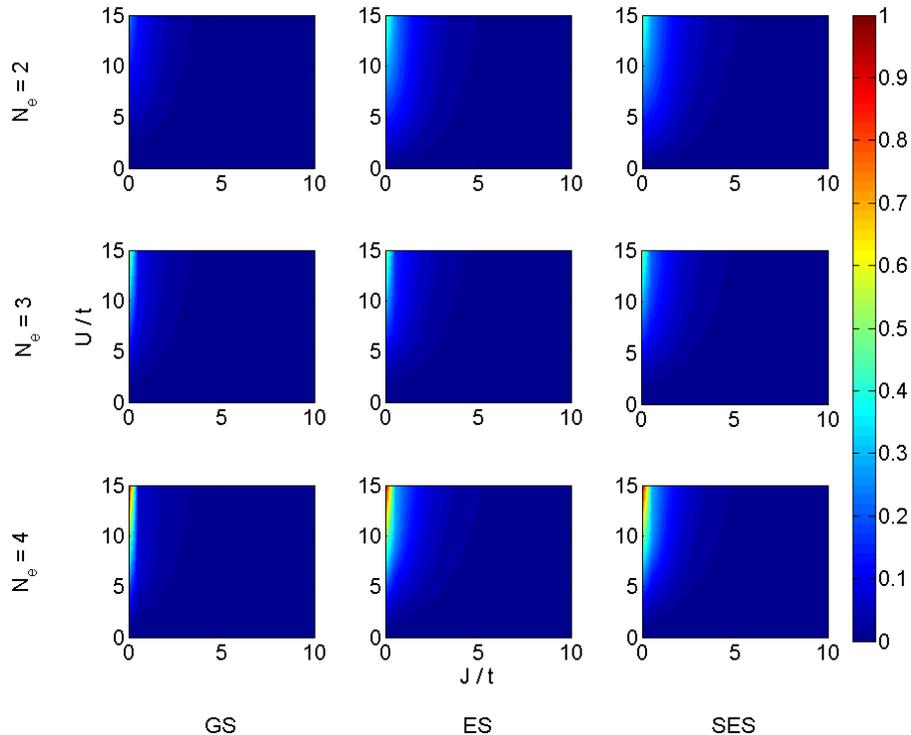} \caption{Relative error between the exact and the Hartree-Fock evaluation of the ground state, first excited state and second excited state energies for an 8 site system, with 2, 3 and 4 electrons, averaged over 25 local moment configurations, as a function of $U/t$ and $J/t$.}
\label{fig:ED} \end{figure}
\end{center}
\end{widetext}

\begin{figure}[htb]
\includegraphics[width=6cm,angle=270]{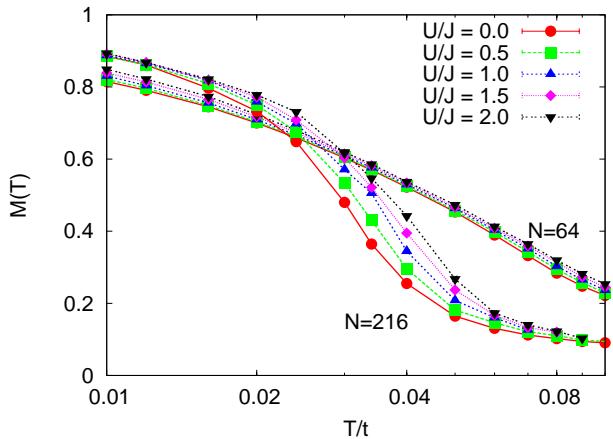}
\caption{Local moment magnetization for $J/t = 5$ for $N = 64$ and 216 local moments 
for $U/J = 0.0$, 0.5, 1.0, 1.5, and 2.0.}
\label{fig:LMJ5}
\end{figure}

We calculated the magnetization of the local moments
$$ M(T) = \left< \frac{1}{NS} \sqrt{\left|\sum_i \bvec{S}_i\right|^2} \right>,$$
as a function of temperature, where the angle brackets indicate a thermodynamic average.
In Fig.~\ref{fig:LMJ5} we show the local moment magnetization when $J/t = 5$ calculated 
both for $N=64$ and $N=216$ local moments.  There is some enhancement of the magnetization with
increasing $U/J$ in the $N=64$ samples.  However for the larger $N=216$ system the enhancement
is much clearer and is more meaningful as 
the $N=216$ data should have smaller finite size effects 
 and be more reflective
of the thermodynamic limit.  Calculations of the fermion magnetization yield 
similar enhancement with increasing interaction strength \cite{Yazdani}.

\begin{figure}[htb]
\includegraphics[width=6cm,angle=270]{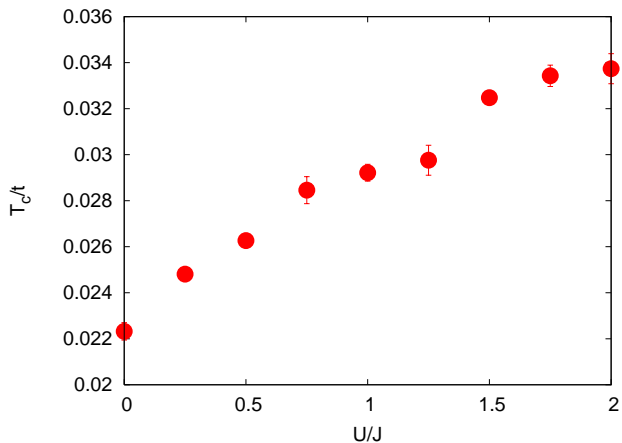}
\caption{Curie temperature as a function of $U/J$ for $J/t = 5$.}
\label{fig:Tc}
\end{figure}

Due to the evident finite size effects in the magnetization, we use the Binder cumulant \cite{Binder} $$g(N,T) = \frac{1}{2}\left[5 - 3 \left(
\frac{\left<M^4\right>}{\left<M^2\right>^2}\right)\right],$$
to determine $T_c$ as a function of
 interaction strength, as $g(N,T)$
should be independent of system size at $T_c$.  Our results for $J/t = 5$, determined using 
$N = 64$ and $N=216$ are displayed in Fig.~\ref{fig:Tc}.  There is a monotonic increase
in $T_c$ with increasing $U/J$, with an almost 50\% increase in $T_c$ between $U/J = 0$
and $U/J = 2$.

By performing Monte Carlo simulations of an interacting DE model, we have 
demonstrated that electron-electron interactions, which are 
known, or expected to be important
in a variety of materials whose magnetic properties are described by DE type models, 
can lead to quantitatively important increases of the Curie temperature. 
 The Hubbard model
can display ferromagnetism \cite{Hubbard}, hence it is natural that in the presence of
Hund coupling, which independently promotes ferromagnetism, the two terms in the Hamiltonian
can reinforce each other to produce the enhancement observed here.
As a mean field approximation, Hartree-Fock will almost certainly overestimate the
tendency to ordering as $U/J$ is increased, but the 
effects observed here should be robust to fluctuations.   This is because 
the interactions enhance an existing tendency 
to ferromagnetism rather than imposing order
on an otherwise disordered system.
Further, our use of Hartree-Fock is confined to 
values of $U/J$ where exact diagonalizations suggest that it will be most accurate.

The scheme we have introduced, and our demonstration that it should be accurate 
for the experimentally relevant interaction range 
$U \lesssim 3J$, should be 
a further step towards the quantitative description of 
the magnetic properties of 
important DE type materials. 
It should also be possible to extend the scheme 
introduced here to treat both disorder 
(which has been included in previous non-interacting simulations 
\cite{Schliemann,Kennett1,Alvarez,Fiete})
and long-range Coulomb interactions, which have been argued 
as being relevant to nanoscale
phase separation and CMR \cite{Shenoy}.

This work was supported by NSERC and most computations were
performed on Westgrid.  We thank Mona Berciu for a critical
reading of a draft of this manuscript and Greg Fiete for 
helpful communications.

\end{document}